\documentclass[12pt, a4paper]{article} 
\usepackage{graphicx} 
\begin{document} 
\noindent 
Empirical Regularities in Distributions of Individual \\Consumption 
Expenditure\\ 
\vspace{0,2cm}\\ 
\noindent 
Martin Hohnisch, Sabine Pittnauer and Manisha Chakrabarty\\ 
Economic Theory II, University of Bonn, D-53113 Bonn, Germany 
\vspace{0,6cm}\\ 
\noindent 
{\footnotesize{ 
We empirically investigate distributions of individual consumption expenditure for four commodity 
categories conditional on fixed income levels. The data stems 
from the Family Expenditure Survey carried out annually in the United Kingdom. We use graphical 
techniques to test for normality and lognormality of these distributions. While mainstream economic theory does not predict 
any structure for these distributions, we find that in at least three commodity categories individual 
consumption expenditure conditional 
on a fixed income level is lognormally distributed. 
\vspace{0,1cm}}} 
\vspace{0,6cm}\\

\noindent 
{\bf{I. Introduction}}\\ 
Probabilistic concepts have been fundamental to economic theories of financial markets for 
almost a century \cite{1} and these markets have received much attention from the econophysics community in recent years 
\cite{2}. With a few exceptions (see \cite{2a}), this has not been the case for other markets, e.g. commodity\footnote{Following economic terminology, 
we use the word commodity for goods and services.} markets although they might deserve broader attention from econophysicists and 
be a fruitful application area for 
methods of computational statistical physics as well.

The prevailing economic framework for describing markets for commodities is 
General Equilibrium Theory \cite{3}. With its main results established about fifty years ago, 
it continues to be a fundamental paradigm in economic thought. 
Its starting point is 
a set $\cal A$ of agents and a space of individual consumption plans $R^{L}_{+}$. Each agent $a \in \cal A$ 
chooses a vector $q \in R^{L}_{+}$ (with coordinate $q_{i}$ denoting the 
quantity of commodity $i$ she wants to consume) as the maximal element with respect to an order relation over $R^{L}_{+}$, 
called her preference or taste, 
subject to the restriction that it must be affordable to her 
at the prevailing price system $p\in R^{L}_{+}$ (with $p_{i}$ denoting the price for commodity $i$). 
Choices of firms regarding supply of commodities are modelled in a similar manner. 
The main success of General Equilibrium Theory has been to prove the existence of a price system $p^{*}$ 
equilibrating aggregate demand and supply on the market for each commodity 
given any specification of preferences and income on the set $\cal A$ 
under very mild 
assumptions on the set $\cal P$ of admissible individual preferences. 
Thus it has established a rigorous framework for understanding how a 
decentralized economic system where individuals and firms decide in a seemingly uncoordinated fashion 
about their individual demands and supplies can become 
self-coordinated by the price system. 
However, General Equilibrium Theory is an intrinsically static concept in many respects and therefore not 
capable of explaining some important issues. 
One major point is that it does not derive endogenously 
the shape and the dynamics of the 
distribution of agents' characteristics, like preference, expectations and income. Because the space of 
admissible distributions of characteristics is not restricted in the model, 
General Equilibrium Theory has too little structure to produce empirically testable predictions 
on market outcomes \cite{4} 
\footnote{The problem lies in the fact, that a distribution of preferences cannot be observed empirically 
because to do so we would need 
to know consumers' consumption decisions for a large set of price systems $p\in R^{L}_{+}$ 
(While relative prices do change over time, so do 
preferences; as a result, a ceteris paribus condition cannot be secured). On the other hand, the distribution of income 
is observable and substantial progress has been achieved in General Equilibrium Theory by taking it into account \cite{5}. 
However, this approach relies on ad-hoc assumptions on the distribution of preferences 
the validity of which cannot be tested within the prevailing theoretical framework.}. 
Already in $1974$, a Markov Random Field model with a finite subset of $\cal P$ as state space has been presented 
\cite{6} 
from which, leaving aside some technical difficulties, distributions of preferences 
can be derived. Unfortunately, this approach received 
little attention in the mainstream economics community despite the contention 
from social sciences 
that interaction between consumers 
is likely to be an important factor determining consumption decisions. 
As a result, to our knowledge no attempts have been made to derive empirically testable predictions from 
probabilistic models of preference dynamics. 
In this paper, we document empirical regularities in the distributions of individual cross-section 
consumption expenditure which might suggest that heterogenity of individual consumption expenditure 
for certain groups of commodities is indeed governed by a common stochastic mechanism. 
\vspace{0,5cm}\\ 
{\bf{II. Data and methodology}}\\ 
Our expenditure data is provided by the Family Expenditure Survey carried out annually since 1957 in the United Kingdom. 
The survey is based on 
a representative sample of about 7000 households which amounts to $0.05\%$ of all households in the United Kindom. 
A household comprises one person living alone or a group of people living at the same address. 
Each household contributes information about its total income and its total expenditure for goods and services 
in a time period of two weeks. Related types of goods and services are grouped into nine categories 
and expenditures are aggregated within each category. 
Information about expenditures is obtained partly by records kept by individual members, partly 
by interview in case of periodic expenditures. 
Details of income are obtained by interview. The periods for the record book and interviews are spread 
evenly over the year. 
Additionally, household characteristics like the number of household members, 
the age of the household head etc. are recorded.
We confine our analysis to the categories Services, Fuel (comprising fuel, light and power), 
Food (comprising food and nonalcoholic 
beverages) and Travel (comprising transport and vehicles). 
For each of these categories, we investigate the distribution of expenditure within an annual sample. 
It is obvious that income is an important determinant of expenditure. 
However, we want to exclude the effect of income heterogenity 
\footnote{The shape and the origin of the income distribution constitute an extremely interesting 
research topic, but it is important to stress that regularities present in the distribution of income are not related 
to the regularities we aim at in this paper.} 
to focus solely on heterogenity of tastes. 
Therefore we aim at estimating the distribution of consumption expenditure for a fixed value of 
income rather than in the whole sample. 
Clearly, we have to base our estimation on subsamples consisting of observations from narrow income intervals. 
Narrowing down these intervals is limited 
by the need to have a sufficient number of points in a subsample. 
Furthermore we include in one subsample only observations from households with a common number of persons, 
because consumption patterns presumably vary with the size of households. 
We choose from each annual sample four subsamples 
with a width of about $0.3\%$ of the total income spectrum. Within each interval we estimated the income distribution 
using nonparametric techniques. Income tends to be spread evenly within these intervals with no notable regularities. 
To eliminate the effects of remaining income variance we corrected individual observations based on 
the slope of the Engel curve which regresses the dependence of consumption on income. With this 
slope being in most cases in the range between -0.1 and 0.3, we found that this procedure has negligible 
effect on the results except for some smoothing. 
Income and consumption expenditure do not correlate in any of the corrected subsample datasets. 
In summary, the stratification procedure 
resulted in 37 subsamples comprising 
22 subsamples with one-person households and 
15 with two-person households 
with the number of observations in each subsample 
between 300 and 700. 
In a first step, we used nonparametric density estimation techniques to get qualitative information on 
the shape of the density functions. In a second step, we used probability plotting \cite{7} to investigate the 
functional type of the distributions. 
In probability plotting, the values of the empirical distribution function are 
transformed in such a way that they will 
follow a straight line if plotted against the observed realizations of the random variable $x$ 
(within sampling error) if the hypothesized distribution is the true 
underlying distribution. 
Assume the true distribution is $F$ with mean $\mu$ and variance $\sigma^2$. We write 
\begin{equation} 
F(x)=G(\frac{x-\mu}{\sigma})=G(z) 
\end{equation} 
If we plot $z=G^{-1}(F(x))=\frac{x-\mu}{\sigma}$ against $x$, the resulting plot will be a straight line. 
Probability plotting displays $z_i=G^{-1}(F_n(x_{(i)}))$ on $x_{(i)}$ 
with the empirical distribution function 
\begin{equation} 
F_n(x_{(i)}) = \frac{i - 0.5}{n} 
\end{equation} 
and the 
ordered observations $x_{(1)}\leq \dots \leq x_{(n)}$. 
If the hypothesized distribution is normal, \cite{7} recommends that 
$F_n(x)$ be transformed by 
\begin{equation} 
z={\rm sign}(F_n(x)-0,5)(1,238t(1+0,0262t)) 
\end{equation} 
with 
\begin{equation} 
t=\{-\ln{[4F_n(x)(1-F_n(x))]}\}^{1/2} 
\end{equation} 
and plotted against $x$. 
If the hypothesized distribution is lognormal, \cite{7} recommends that the same transformation be applied on $F_n(x)$ and 
$z$ to be plotted against $\ln x$. 
\vspace{0,5cm}\\ 
{\bf{III. Results}}\\ 
Nonparametric estimates show that for each category and in all subsamples the distribution of 
consumption expenditure is unimodal. The estimated density function oscillates in the tails due to limited 
sample size. The magnitude of these oscillations is similar as in nonparametric density estimates of 
Monte-Carlo generated samples from lognormal 
distributions presented in the literature \cite{8}. 
For the good categories Services, Fuel and Travel, the nonparametric density estimates indicate that 
the distributions of expenditure are skewed to the right 
(see top of Figures 1, 2, 3 for representative plots). 
In the category Food, 
the estimated distributions appear to be 
slightly skewed to the right.
With these preliminary findings, we tested the data of each subsample and for each category for normality and lognormality 
using probability plotting. 
In the categories Services, Fuel and Travel, the values obtained by formulae 
(3) and (4) follow a straight line in 
lognormal probability plots within sampling error indicating lognormality of the distributions 
(see bottom of Figures 1, 2, 3 for representative plots). 
In a few plots there are outliers present in the upper and lower ends of the distribution which appear to result from 
contamination. Based on \cite{9}, where Monte-Carlo generated samples from mixtures 
of two lognormal distributions are displayed in probability plotting, we concluded that the weight of a possible 
contaminant distribution is less than 0.2. 
For the category Food, it is difficult to distinguish between normality and lognormality 
from probability plotting. In normal probability plotting, we obtain in most subsamples a slightly 
concave curve 
indicating that the distribution is skewed to the right \cite{7} while obtaining 
a straight line in the remaining cases indicating normality. 
In the former instance, the sample points 
follow a slightly convex curve in lognormal probability plotting indicating a deviation from lognormality 
towards normality \cite{7}.
\begin{figure}[hbt] 
\begin{center} 
\includegraphics[angle=-90,scale=0.5]{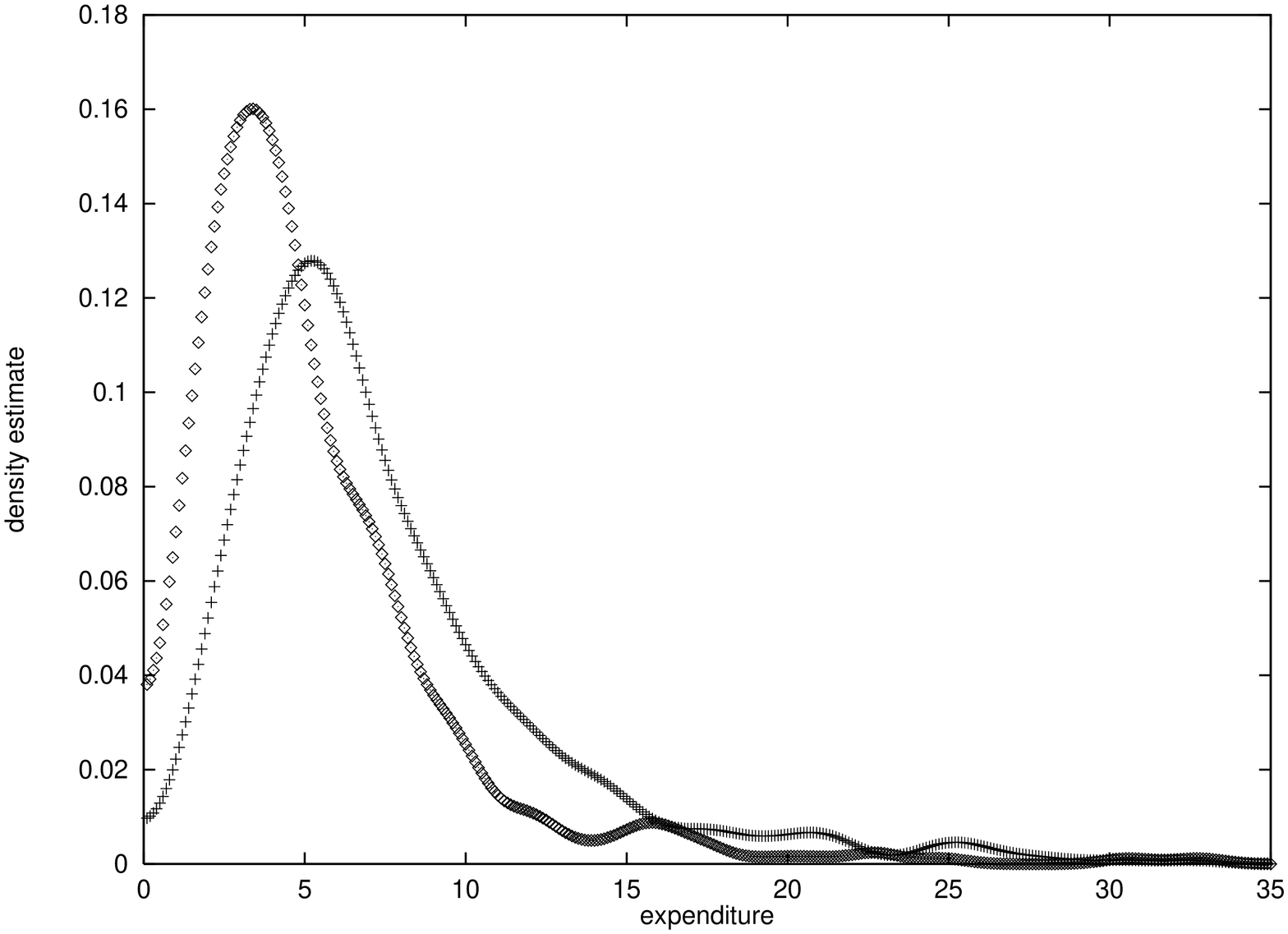} 
\includegraphics[angle=-90,scale=0.5]{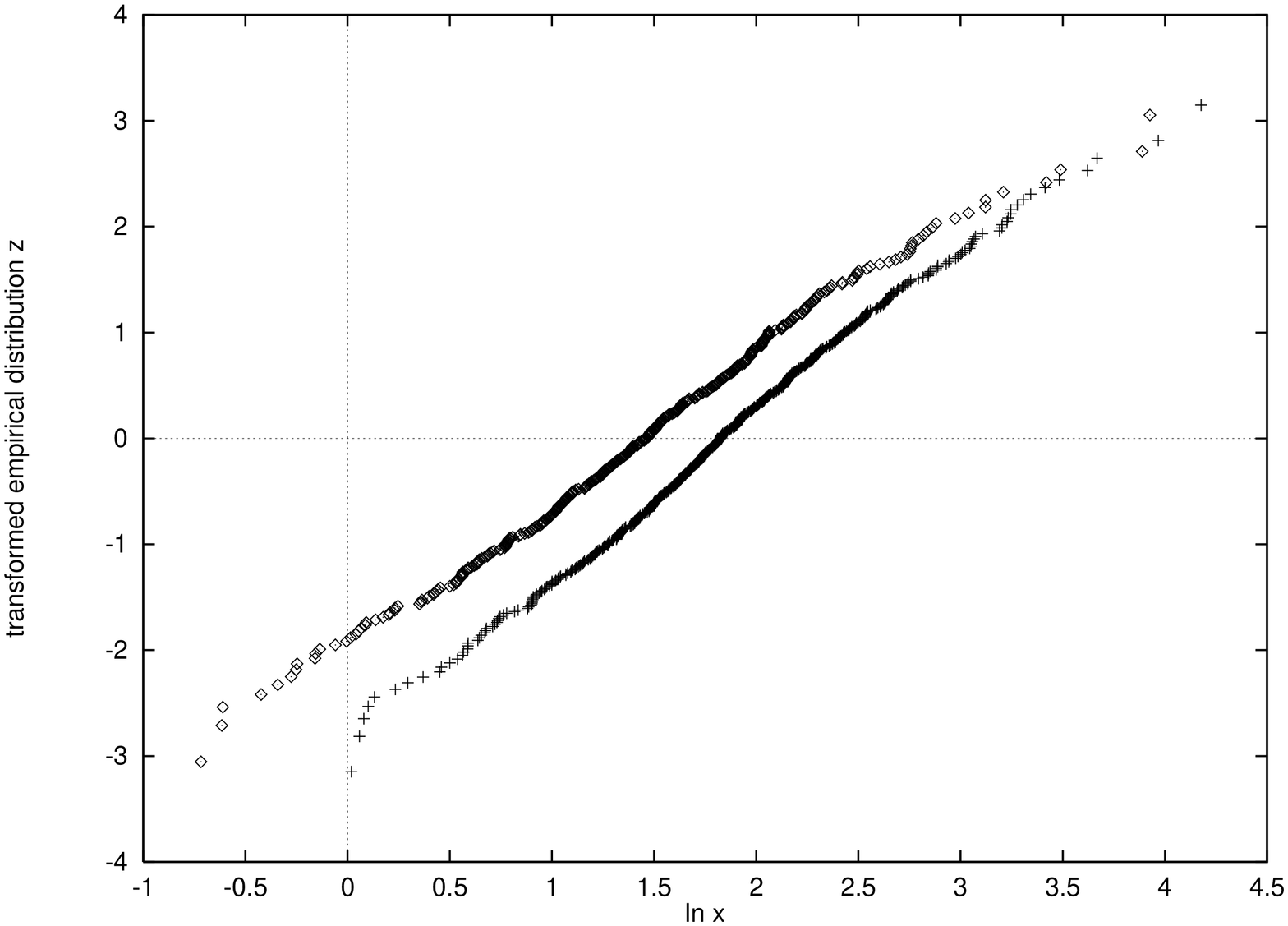} 
\end{center} 
\caption{ 
Representative plots for the category Services; top: 
nonparametric density estimates for the subsamples: 1987, 1 person, income 40-70 (diamonds)
 and 1986, 2 persons, income 100-150 (+); 
bottom: lognormal probability plots for the same subsamples 
} 
\end{figure}
\begin{figure}[hbt] 
\begin{center} 
\includegraphics[angle=-90,scale=0.5]{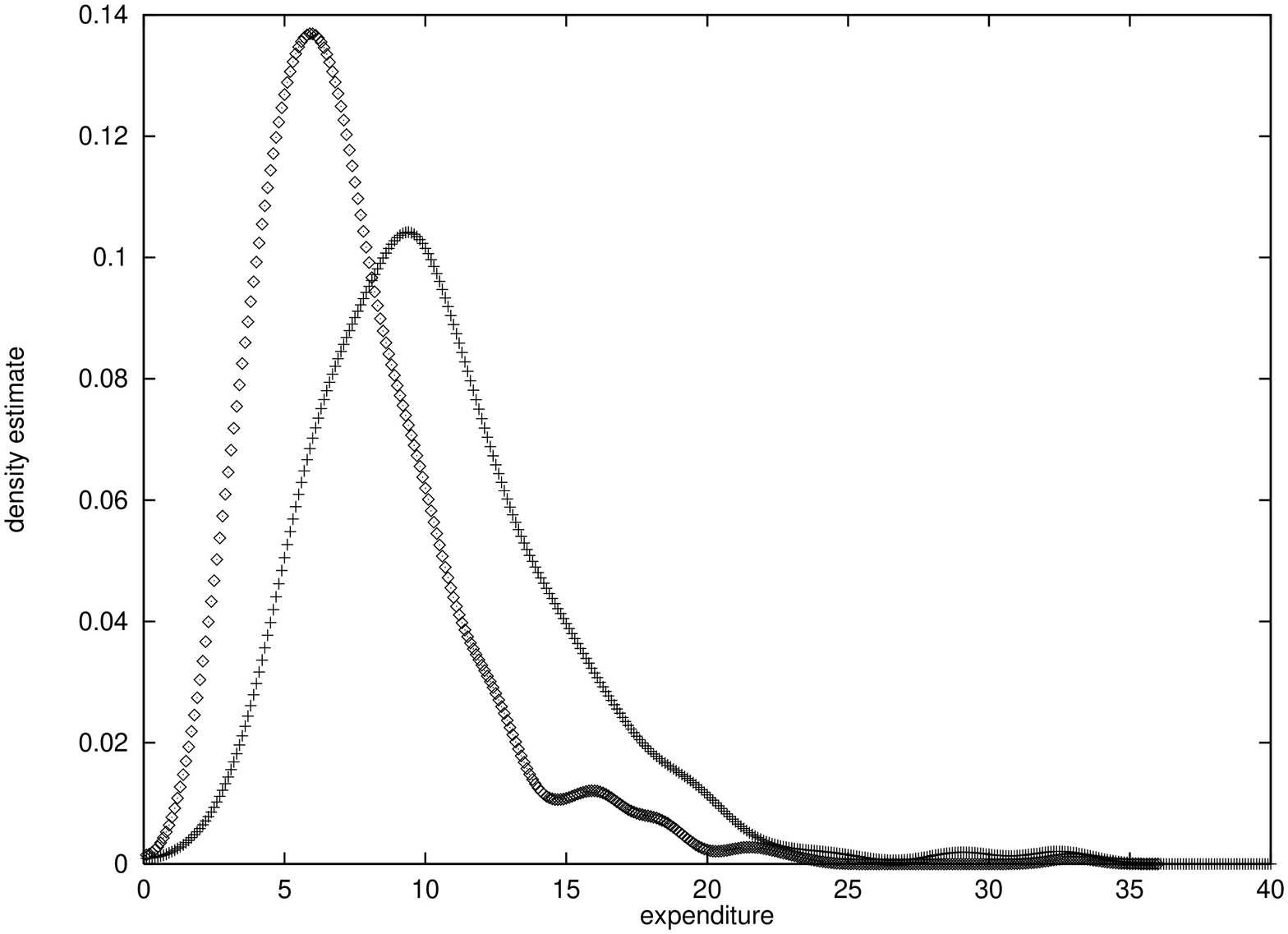} 
\includegraphics[angle=-90,scale=0.5]{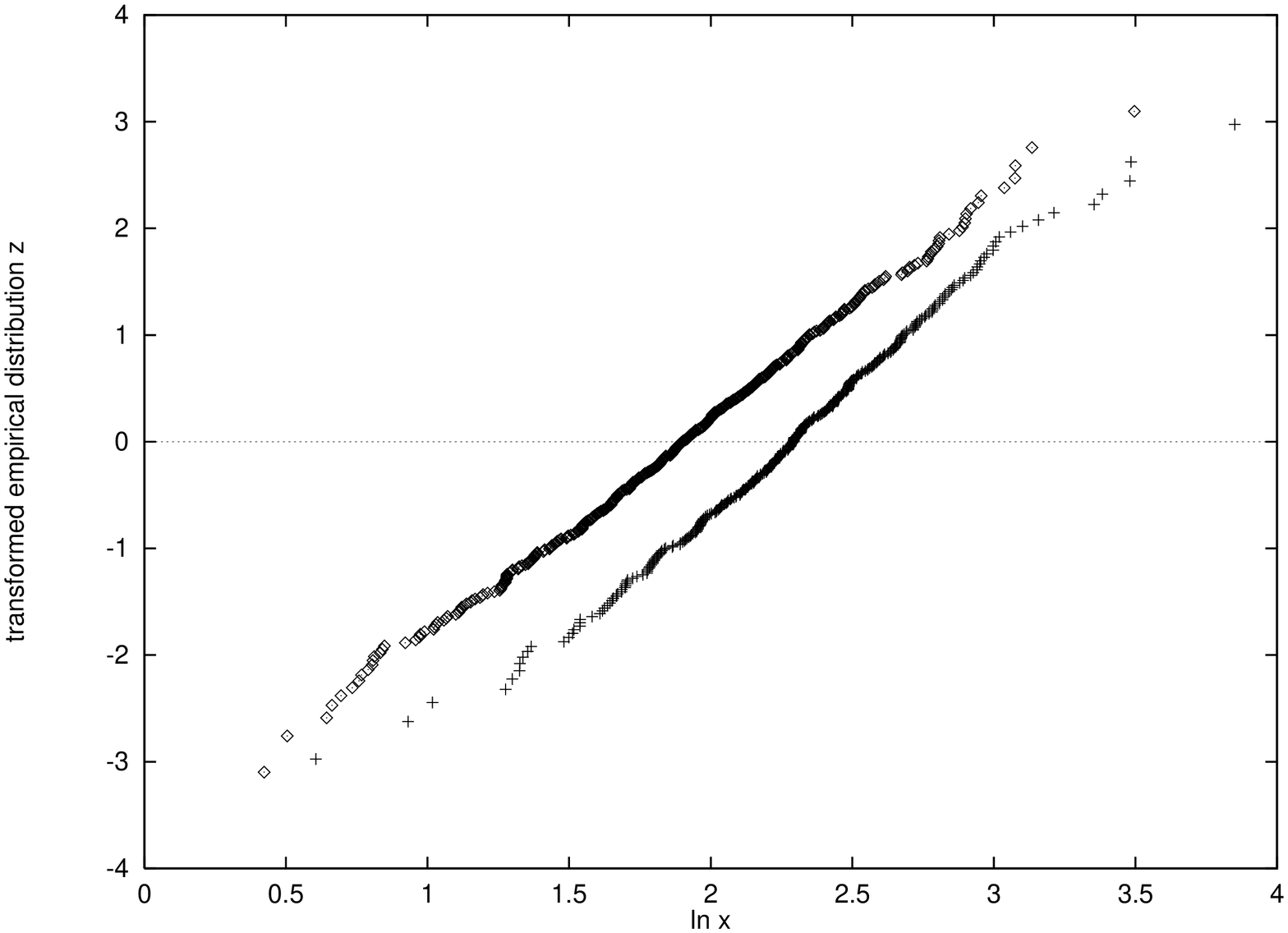} 
\end{center} 
\caption{ 
Representative plots for the category Fuel; top: 
nonparametric density estimates for the subsamples: 1988, 1 person, income 70-100 (diamonds)
 and 1992, 2 persons, income 200-250 (+); 
bottom: lognormal probability plots for the same subsamples 
} 
\end{figure}
\begin{figure}[hbt] 
\begin{center} 
\includegraphics[angle=-90,scale=0.5]{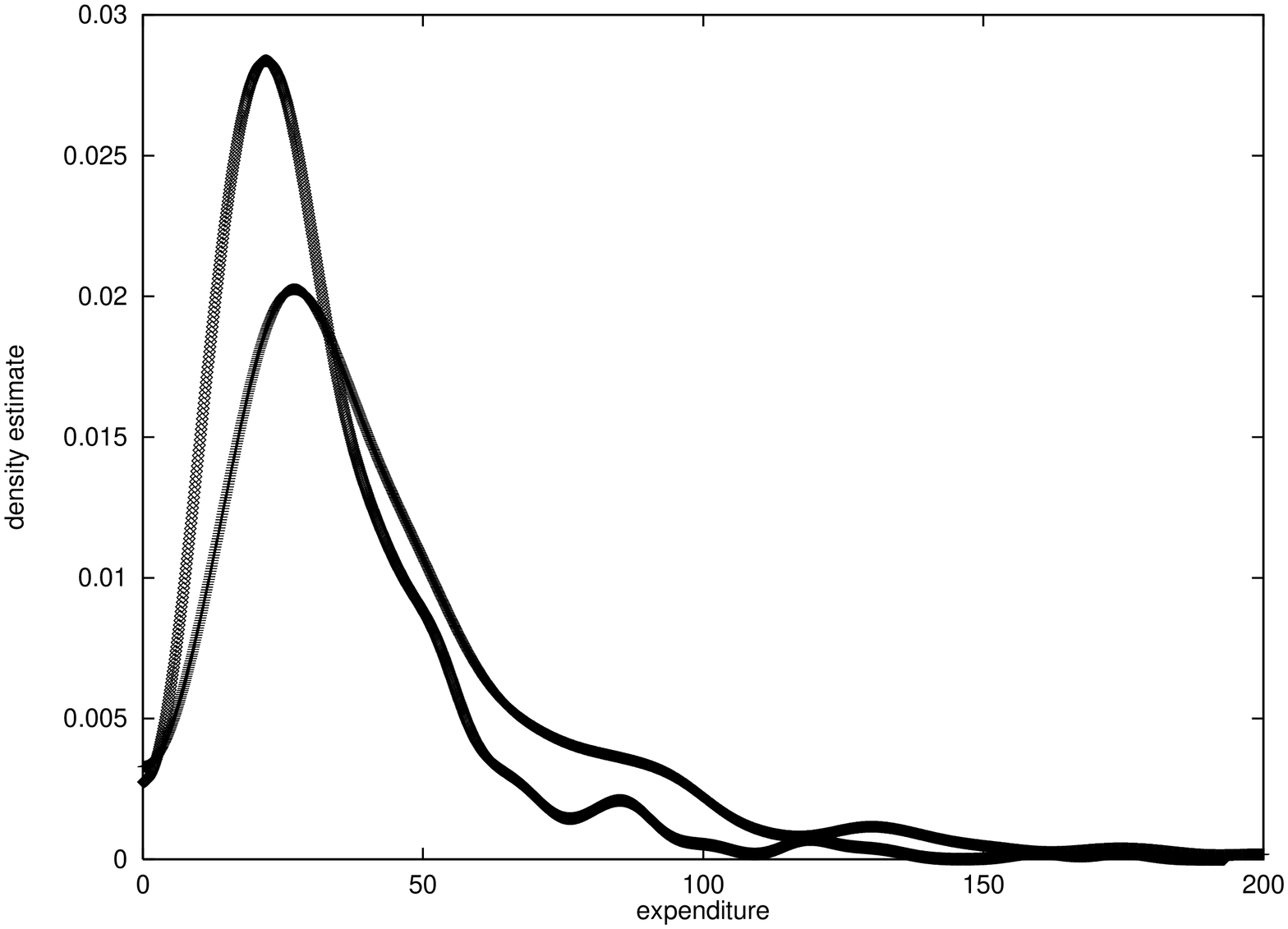} 
\includegraphics[angle=-90,scale=0.5]{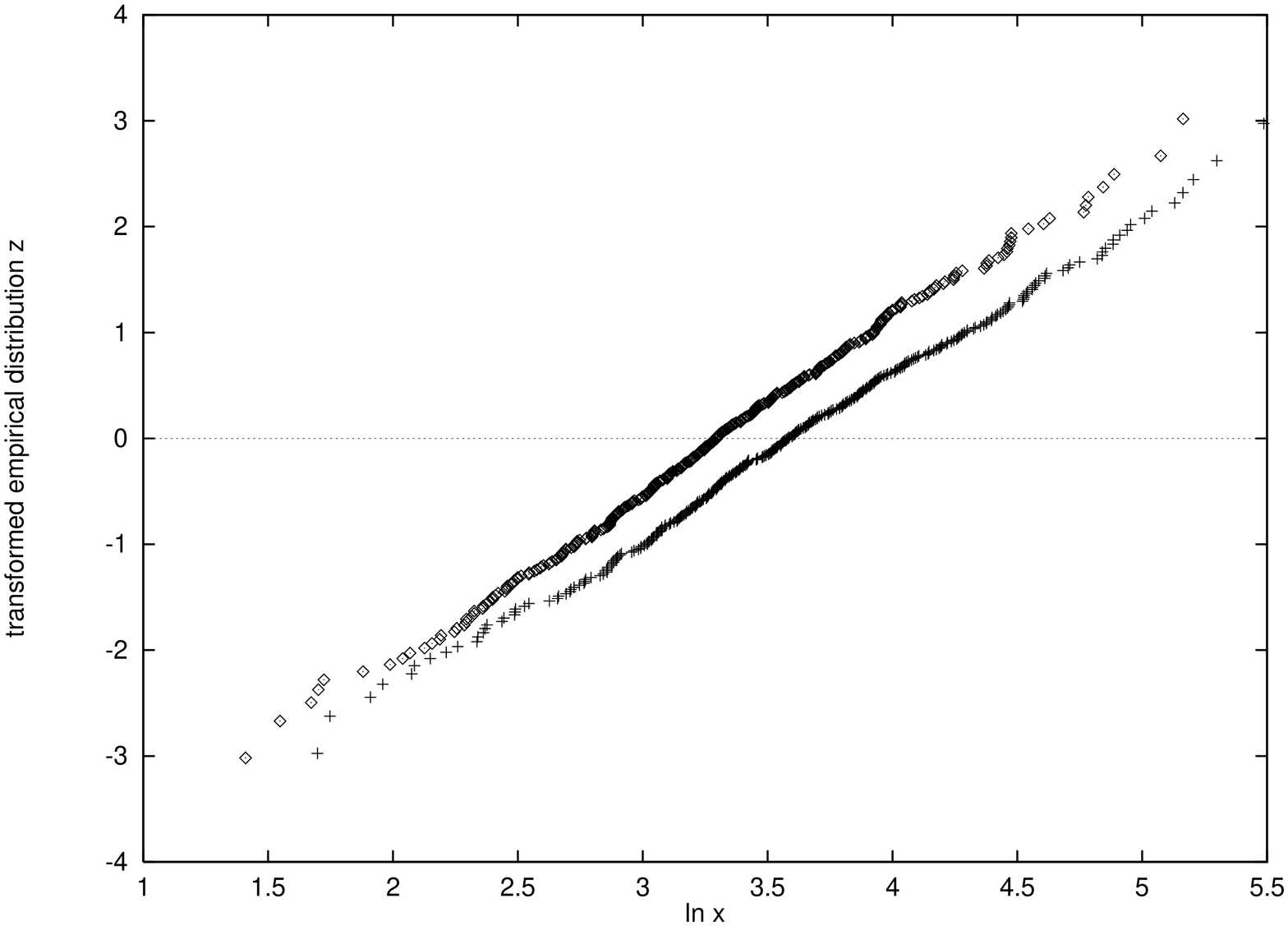} 
\end{center} 
\caption{ 
Representative plots for the category Travel; top: 
nonparametric density estimates for the subsamples: 1988, 2 persons, income 150-200 (diamonds) 
and 1992, 2 persons, income 200-250 (+); 
bottom: lognormal probability plots for the same subsamples 
} 
\end{figure}
\noindent
\vspace{0,5cm}\\ 
{\bf{IV. Discussion}}\\ 
We see two potential explanations for the regularities found in this paper. 
First, the distributions might originate simply from fluctuations inherent in the process by which the data is 
obtained. 
The reported individual expenditures 
within a given category involve adding amounts from many instances of trading 
for goods and services of many types and brands. 
However, the fact that in at least three good categories we find lognormal distributions 
makes it unlikely that the regularities are attributable solely 
to random fluctuations. 
By the Central Limit Theorem a lognormal distribution of an observable would originate from 
a multiplicative process involving stochastically independent fluctuation on each stage, 
but we do not see how a multiplicative process could 
be involved in the process by which our data is obtained.
Therefore we suggest that the observed regularities have a second - and deeper - origin in a 
stochastic process governing the 
heterogenity of individual tastes \footnote{In economic theory, 
heterogenity of tastes would be formalised as 
a distribution on an infinite dimensional space of functions representing preferences, as outlined 
in the introduction. However, the notion of preferences is a hypothetical concept and one might well doubt 
their existence if this concept does not provide testable predictions.}. 
Lognormal distributions are ubiquitous in natural sciences where their origin is some structure of the underlying system. 
The question of whether the regularities in consumption data have some deeper origin lying in the structure of 
socioeconomic systems is presumably worth exploring.
\vspace{0,5cm}\\ 
The authors thank J. Arns for extracting the data from the Family 
Expenditure Survey. We are indebted to W. Hildenbrand and D. Stauffer for 
many insightful discussions. S.P. gratefully acknowledges financial support 
from the Graduiertenf\"orderung, University of Bonn.

\end{document}